# ИНФОРМАТИКА

УДК 004.9:66.013.512

## МОДЕЛИРОВАНИЕ СТРОИТЕЛЬНОЙ ПОДОСНОВЫ В САПР РЕКОНСТРУКЦИИ ПРЕДПРИЯТИЙ НА ОСНОВЕ МОДУЛЕЙ В ЧЕРТЕЖЕ

В.В. Мигунов

*ЦЭСИ РТ при КМ РТ, г. Казань*

Аннотация

Описаны параметрическая модель, состав и особенности реализации операций разработки чертежей строительной подосновы - общей составляющей чертежей различных марок в проектах реконструкции предприятий. Ключевым моментом углубленной автоматизации проектирования явилось применение так называемых модулей в чертеже, объединяющих видимую графическую часть и невидимые параметры. Модель прошла проверку при подготовке нескольких сот чертежей.

Библ.4

Abstract

V.V. Migunov. The modelling of the build constructions in a CAD of the renovation of the enterprises by means of units in the drawings // Izvestiya of the Tula State University/ Ser. Mathematics. Mechanics. Informatics. Tula: TSU, 2004. V._. N _. P. __−__.

The parametric model of build constructions and features of design operations are described for making drawings, which are the common component of the different parts of the projects of renovation of enterprises. The key moment of the deep design automation is the using of so-called units in the drawings, which are joining a visible graphic part and invisible parameters. The model has passed check during designing of several hundreds of drawings.

Bibl.4

Настоящая работа посвящена применению модульной технологии разработки проблемно-ориентированных расширений систем автоматизированного проектирования (САПР) реконструкции предприятия, общие положения которой изложены в [1]. Основой этой технологии являются модули в чертеже - дуальные объекты, включающее видимую геометрическую часть и невидимое параметрическое представление (ПП). Первично ПП, по которому генерируется видимая часть.

Объект приложения технологии - автоматизация подготовки чертежей так называемой строительной подосновы. Подоснова включает координационные оси зданий и сооружений, колонны, перегородки, проемы и другие строительные конструкции, отражаемые в чертежах различных марок системы проектной документации для строительства (СПДС) согласно [2]. К элементам строительной подосновы, в частности, осуществляется привязка при монтаже оборудования, технологических трубопроводов, электроснабжения, внутренних сетей водоснабжения и канализации, средств автоматизации и т.д. в ходе реконструкции предприятий. В СПДС на изображении каждого здания или сооружения указывают координационные оси и присваивают им самостоятельную систему обозначений. Это касается не только проектной и рабочей документации на строительство предприятий, зданий и сооружений различного назначения, но и отчетной технической документации по инженерным изысканиям для строительства [2].

Один и тот же чертеж строительной подосновы используется многократно в чертежах нескольких марок, выполняемых специалистами различных профилей в рамках одного проекта. Задачи реконструкции предприятий требуют повторного изображения неизменной части строительной подосновы в разных проектах, выполняемых в различное время, последовательно. Тем самым эффективность автоматизации проектирования подосновы резко повышается по сравнению с автоматизацией других частей проекта, которые реже используются повторно.

Согласно модульной технологии разработки расширений САПР, после выяснения целесообразности разработки специализированного расширения необходимо выявить наиболее информационно связанные элементы чертежа, изображения которых имеет смысл генерировать автоматически по ПП.

Как показывает анализ чертежей различных марок и требований ГОСТов СПДС к ним, наибольшей связанностью обладают координационные оси, колонны, перегородки (стены), проемы, тексты. Чаще всего используются поэтажные планы (рис.1). Для чертежей марок КЖ (конструкции железобетонные), КМ (конструкции металлические), АС (архитектурно-строительные решения) также важны элементы фундаментов и перекрытий/покрытий, иногда требуются и разрезы с отметками высоты. Из-за сильной связанности с поэтажными планами фундаментов (башмаки идут под колонны, балки опираются на башмаки, ленточные фундаменты устанавливаются под перегородки) и перекрытий/покрытий (балки опираются на колонны, плиты опираются на балки) целесообразно автоматизировать проектирование всех перечисленных объектов совместно. В таблице 1 приведен состав учитываемых объектов чертежа для трех видов чертежей в плане.

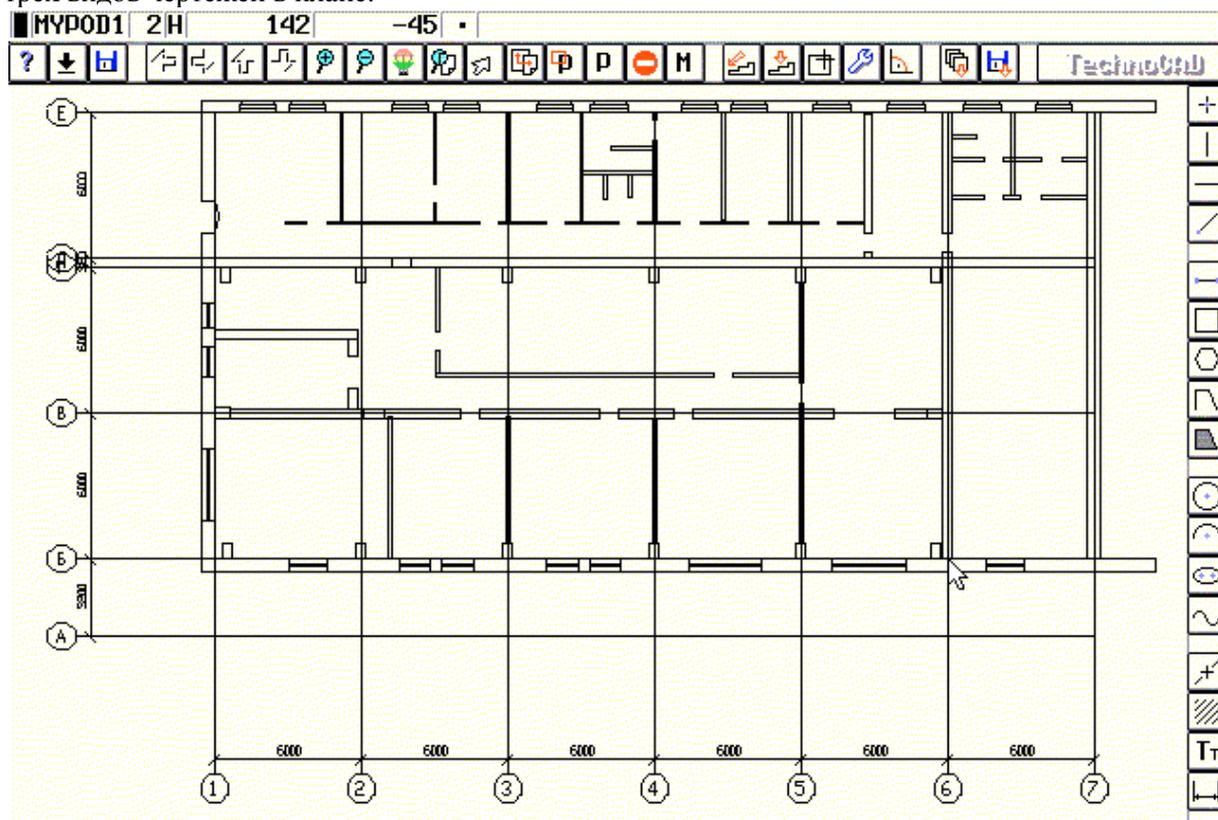

Рис.1. Пример плана этажа

Таблица 1. Объекты в планах этажа, покрытия/перекрытия и фундамента

| Список объектов | Этаж | Покрытие/перекрытие | Фундамент |
| --- | --- | --- | --- |
| группы горизонтальных осей | + | + | + |
| группы вертикальных осей | + | + | + |
| группы колонн | + | + | – |
| перегородки (стены) | + | + | – |
| проемы | + | – | – |
| балки перекрытия/покрытия | – | + | – |
| группы плит | – | + | – |
| ленточные фундаменты | – | – | + |
| группы башмаков | – | – | + |
| фундаментные балки | – | – | + |
| тексты | + | + | + |

В таблице отсутствуют такие элементы чертежа подосновы, как размеры, обозначения осей, отметки высоты. Их генерация производится по общим установкам и свойствам объектов, вошедших в таблицу. Такой подход реализован в САПР TechnoCAD GlassX [3]. Количественные сведения и примеры, приводимые ниже, относятся также к этой системе.

Параметрическое представление объектов и их связей в чертеже строительной подосновы содержит списки объектов, в описании которых отражены связи принадлежности (ссылки по номеру объекта в своем списке). Точка пересечения двух координационных осей задается номерами этих осей в общей нумерации и называется далее узлом привязки. От таких узлов отсчитываются смещения колонн, перегородок, проемов и др. Все координаты, размеры и смещения задаются в миллиметрах натуры, если не указано иначе. Заданный узел привязки и смещение от него начала объекта (левого нижнего угла) далее называются привязками. Признак новизны для объекта означает, существующий он или новый (существующие вычерчиваются тонкими линиями). Признак X задает, что элемент ориентирован вдоль оси X, и вдоль нее отсчитывается его длина, а по нормали - ширина; иначе наоборот. Имеются следующие списки объектов.

*Группы горизонтальных осей.* Группы горизонтальных координационных осей задаются числом осей в группе (до 99), признаком основные/дополнительные оси, расстоянием до следующей оси в случае основных осей либо смещением от основной для дополнительных.

*Группы вертикальных осей.* Имеют те же свойства, что и группы горизонтальных осей.

*Группы колонн.* Марка колонн определяет их ширину и толщину, наличие симметрии, количество и направления балок, которые можно оперетъ на колонну, и др. Имеется более 600 вариантов марок. Например:
- "3К96-7      (10500 х  600 х 400, 2.300)" - "Колонны одноэтажных зданий с мостовыми кранами. Серия 1.424.1-75". "Колонны крайних рядов. Выпуск 1/87, 2/87.";
- "Немаркированная колонна" - для немаркированных колонн задается длина консоли или ветви, а также тип колонны, например: "Железобетонная двухконсольная", "Металлическая двухветвевая".

Задаются узлы привязки начала и конца группы колонн, смещение центров колонн от узлов привязки, признак X, признак новизны. Для одноконсольных колонн задается признак расположения консоли слева (снизу).

*Перегородки (стены).* Характеризуются типом по ГОСТ 21.107-78: обыкновенная, сборная щитовая, из стеклоблоков, остекленная 1 (три продольных линии), остекленная 2 (четыре продольных линии), кирпичная. Задаются толщина и длина перегородки, признак несущая/не несущая, признак X, привязка, признак новизны.

*Проемы.* Характеризуются маркой (около 100 вариантов). Например:
- "ОР 15-6      (1460 х  570)" - окно с двойным остеклением для жилых и общественных зданий по ГОСТ 11214-86;
- "ДН 21-13АПЩ (2085 х 1274, АПЩР2)" - дверь наружная для жилых и общественных зданий по ГОСТ 24698-81;
- "Немаркированный проем" - если марки нет

Тип проема по ГОСТ 21.107-78 имеет 19 вариантов. Например:
- "Проем без четвертей (не доходящий до пола)";
- "Дверь складчатая в проеме без четвертей".

Задаются ширина и высота проема, ссылка на перегородку, в которой выполнен проем, признак X, признак поворота проема на 180 градусов, привязка, признак новизны.

Если планируется последующая генерация разрезов подосновы, задаются дополнительные сведения о проеме: высота нахождения проема над уровнем пола, собственно высота проема, признак наличия, марка, длина, ширина и высота перемычки. Вариантов марок перемычек - более 60. Например:
- "2ПБ19-3-п (1940 х 120 х 140, 0.033)" - Перемычки брусковые. Серия 1.038.1-1, вып.1;
- "2ПП18-5  (1810 х 380 х 140, 0.096)" - Перемычки плитные. Серия 1.038.1-1, вып.2,5.

Также для разрезов задаются признак наличия, марка, толщина, ширина и высота фрамуги. Имеется 10 вариантов марок фрамуг. Например:
- "ФВ 04-12  ( 390 х 1170)";
- "ФВ 13-10  (1290 х  970)".

*Балки перекрытия/покрытия.* Марка балки выбирается из 140 вариантов. Например:
- "2БСО 12-6 АШв   (11960 х 280 х  890, 2.00)" - "Балки стропильные. Серия 1.462.1-1/88  вып.1";
- "ИБ 8-21       ( 5280 х 800 х  300, 1.23)" - "Ригели производственных зданий. Серия ИИ 23-3.70".

Задаются длина, ширина и высота балки, привязка, признак X, признак новизны, привязка левого (нижнего) конца балки к колонне (номер группы колонн, номера внутри группы колонн вдоль осей X и Y) и аналогичная привязка правого (верхнего) конца балки.

*Группы плит.* Марка плиты выбирается из 180 вариантов. Например:
- "2ПВ12-5-4      (11960 х 2980 х 525, 3.200)" - "Плиты покрытия ребристые. Серии 1.465-3, 1.465.1-3, 1.465.1-7";
- "ПК24.12-8Т    (2380 х 1190 х 220,  0.35)" - "Плиты многопустотные. Серия 1.141-1, выпуск 60".

Задаются длина, ширина и высота плиты, признак X, привязка, число плит в группе.

*Ленточные фундаменты.* Задаются: ширина и длина, признак X, привязка, признак новизны.

*Группы башмаков.* Марка башмака выбирается из 25 вариантов. Например:
- "1Ф 12.8-1 (1200 х 1200 х  750, 0.75)";
- "2Ф 18.9-2 (1800 х 1800 х  900, 1.60)".

Задаются длина, ширина и высота башмака, признак X, узлы привязки начала и конца группы башмаков, смещение центров башмаков от узлов привязки, признак новизны.

*Фундаментные балки.* Марка балки выбирается из 70 вариантов. Например:

- "1БФ6-5      (5050 х 200 х 300, 0.27)" - "Серия 1.415.1-2, выпуск 1. ";
- "ФБ 6-36     (5050 х 450 х 520, 0.75)" - "Серия 1.415-1, выпуск 1. ФБ";
- "Немаркированная".

Задаются длина, ширина и высота балки, привязка, признак Х, признак новизны, привязка левого (нижнего) конца балки к башмаку (номер группы башмаков, номера внутри группы башмаков вдоль осей X и Y, положение балки на башмаке: по центру, по левому или правому краю), привязка правого (верхнего) конца балки к башмаку (номер группы башмаков, номера внутри группы башмаков вдоль осей X и Y).

*Тексты.* Сам многострочный текст с установками шрифта, шага строк и др. - как у обычного текста в чертеже. Всегда имеет сноску. Задаются также точка начала текста и точка указания сноски.

Совокупность списков объектов – реляционная база данных с поддержкой ссылочной целостности. Кроме списков объектов, в ПП входят общие для всей строительной подосновы параметры - установки, такие, как смещение наименований осей и размеров от крайних осей; признак того, что горизонтальные оси нумеруются буквами, а вертикальные цифрами, а не наоборот; признак расположения горизонтальных размеров сверху от осей и другие.

Специализированные структуры данных в ПП и, прежде всего, связи объектов по ссылкам, сильно повышают возможную степень автоматизации проектных работ. Например, при изменении шага в группе координационных осей автоматически двигаются связанные с ними колонны, перегородки, проемы на перегородках, перегенерируются тексты размеров. При выборе марок строительных конструкций из имеющихся вариантов автоматически определяется часть их размеров, они заданы в скобках в наименованиях марок. ПП может записываться на диск (без геометрических элементов), порождая информационную среду проектирования в виде библиотек прототипов. При выборе ПП для чтения с диска геометрия генерируется в режиме on-line, и проектировщик легко ориентируется в прототипах. Хранение ПП на диске и в модулях в чертеже компактно - для чертежа рис.1 это около 2 килобайт. Оси строительной подосновы можно автоматически вставить в модуль аксонометрической схемы [4], выбрав модуль подосновы в чертеже или ПП на диске.

Все работы по модификации ПП строительной подосновы производятся в специальном режиме, в собственном основном меню. Используется как пользовательский интерфейс общего назначения (меню, формы ввода...), так и специализированный для черчения и корректировки ("чертежный"), к которому пользователь привыкает во время работы с самим чертежом. В основе реализации "чертежного" интерфейса пользователя лежат специальные временные модули. Они помещаются в чертеж только на время работы в этом специальном режиме и позволяют организовать работу методами, уже развитыми в САПР для других элементов чертежа.

Каждый рабочий модуль соответствует одному объекту из вышеназванных списков. Модуль содержит его изображение, но не включает его свойств, хранимых в ПП. Временный модуль содержит ссылку на свой объект: идентификатор списка объектов и номер объекта в этом списке. Таким образом, выбор в чертеже временного модуля (модулей) эквивалентен выбору объекта (объектов) в параметрическом представлении. Поскольку при модификациях ПП ссылки на объект могут стать некорректными, при всяком изменении ПП осуществляется частичная или полная перегенерация временных модулей. На основе применения временных модулей удается обеспечить углубленную автоматизацию разработки чертежей строительной подосновы, автоматизируемые операции которой характеризуются ниже.

Проектировщик избавляется от необходимости вычерчивания повторяющихся элементов, выполняя только неизбежные функции принятия решений. Максимально используется перегенерация изображения при перемещении курсором характерных точек.

При создании всех видов планов подосновы автоматизированы следующие этапы работ:
автоматически проставляются размеры пролетов;
при необходимости можно сгенерировать общий размер всех пролетов или проставить наименования осей и размеры с другой стороны от плана;
задается размер шрифта, которым будут выводиться все генерируемые размеры и отметки высот (при генерации разреза);
извлечение параметров существующего плана, как находящегося в чертеже, так и на диске в ПП.
При создании плана этажа:
маркировка осей задается указанием начальных букв и цифр;
оси разбиваются на группы, каждая из которых характеризуется шагом осей (длиной пролетов). Нанесение и правка групп X и Y осей происходит путем указания на группу осей в чертеже, вводом значения шага осей и количества осей в группе. Количество осей в группе можно задать построением, указав в чертеже положение крайней оси. При перемещениях курсора прорисовываются оси строящейся группы с заданным шагом, и показывается их количество (рис. 2, 3). В отличие от планов фундамента и перекрытия/покрытия группы осей можно удалять, добавлять, изменять в любой момент, а не только в случае отсутствия других элементов;

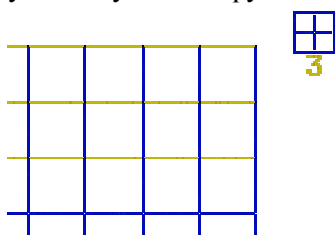

Рис.2. Указание количества горизонтальных осей с подсветкой числа пролетов. Серый цвет – оси, появляющиеся и исчезающие при перемещении курсора

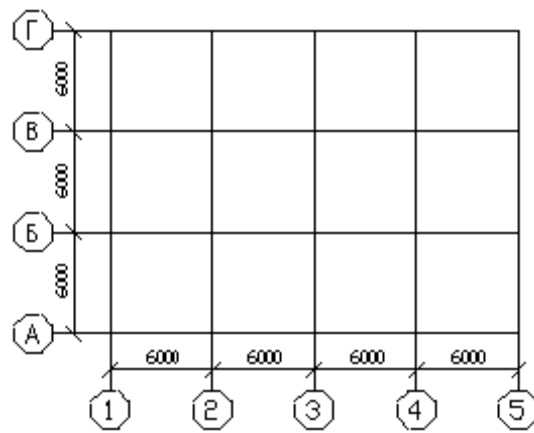

Рис.3. Координационные оси, сгенерированные по указанию рис.2

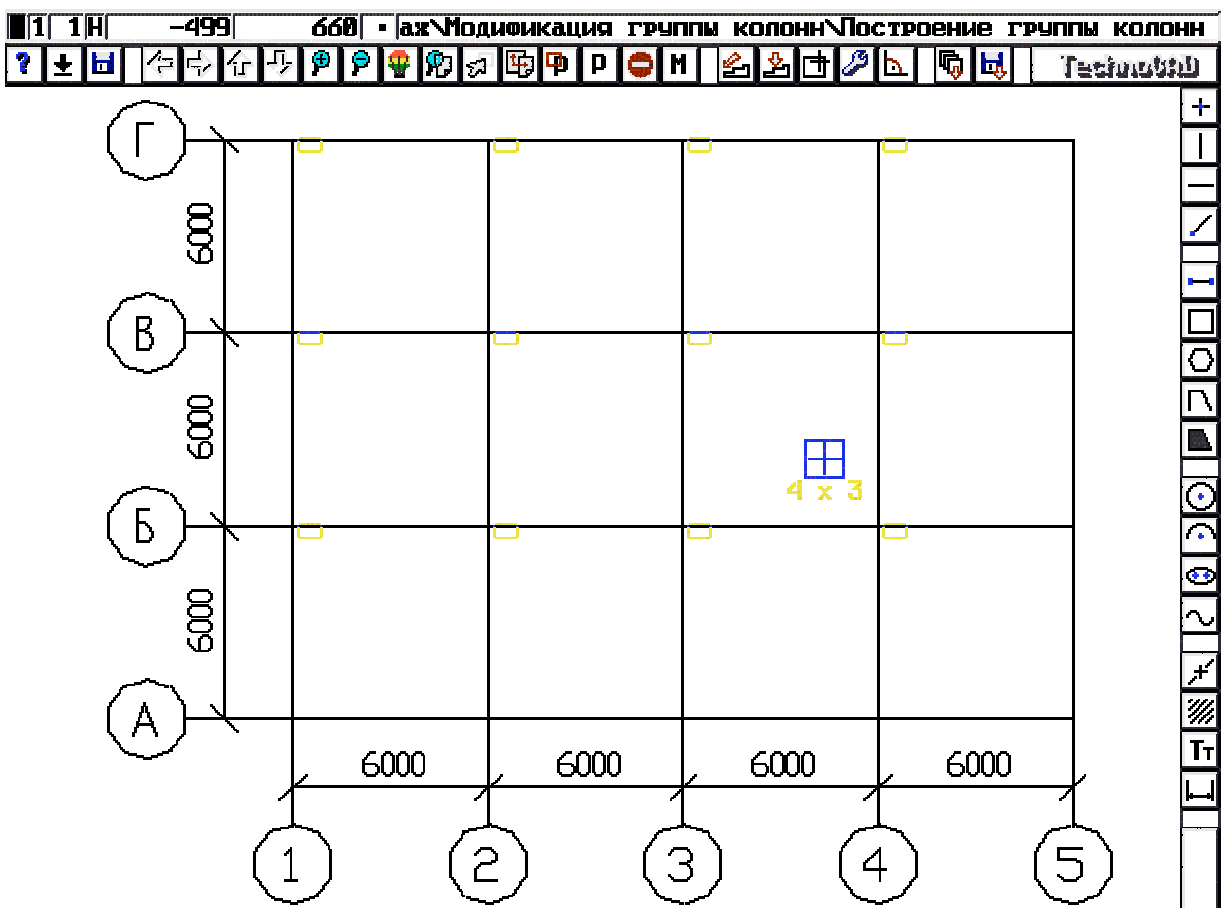

Рис.4. Построение группы колонн. Серый цвет - строящаяся часть нанесение групп колонн осуществляется путем выбора в меню нужной марки или типа (для немаркированных) колонн, указанием в чертеже положений крайних колонн (при перемещениях курсора прорисовываются колонны строящейся группы, и подсвечивается их число, рис.4) и затем уточнением их параметров (для каждого типа – своих). Группы колонн можно удалять, добавлять, корректировать – чертеж автоматически перегенерируется;

нанесение перегородок производится путем указания параметров перегородки и построения ломаной – базовой линии перегородки. Во время построения текущий вид такой составной перегородки виден на экране. Автоматически контролируется расположение перегородок вдоль осей X или Y. Перегородки можно "образмерить", т.е. будут проставлены размеры всех проемов, расположенных на перегородке и размеры всех частей перегородки, находящихся между проемами;

проемы наносятся на перегородки путем выбора нужных марки (проем может быть немаркированным) и типа проема, указания длины и высоты для немаркированного проема. Проект проема появляется на курсоре и движется вместе с ним, при нахождении вблизи перегородки проем "вписывается" в нее автоматически. Если возможны варианты установки проема (дверь внутрь или дверь наружу и др.), они переключаются клавишей. Автоматически контролируется невыход проема за перегородку и непопадание на другой проем. Проемы можно копировать, переносить, удалять, менять их марку, тип и другие параметры (в том числе наличие перемычки и фрамуги и их марки) – чертеж автоматически перегенерируется.

При создании плана фундамента:

новый план фундамента создается на основе плана этажа. При этом из плана этажа импортируются оси, группам колонн сопоставляются группы башмаков, а перегородкам ленточные фундаменты;

группы башмаков можно удалять и корректировать – чертеж автоматически перегенерируется;

нанесение ленточных фундаментов производится путем указания параметров и построения ломаной - базовой линии фундамента. Во время построения текущий вид такого составного ленточного фундамента виден на экране;

нанесение фундаментных балок производится путем указания параметров балки, в том числе ее положения на башмаке (по левому краю, по центру или по правому краю). Затем выбираются башмаки, на которых будет лежать балка. Возможность положить балку на башмак, расположение балки вдоль осей X или Y и соответствие длины балки расстоянию между башмаками контролируется автоматически. Фундаментные балки можно удалять и копировать;

При создании плана перекрытия/покрытия:

новый план перекрытия/покрытия создается на основе плана этажа. При этом из плана этажа импортируются оси, группы маркированных колонн, на которые можно класть балки и несущие перегородки;

нанесение балок производится путем указания параметров балки и выбором колонн, на которых будет лежать балка. Возможность положить балку на колонну в заданном направлении, расположение балки вдоль осей X или Y и соответствие длины балки расстоянию между колоннами контролируется автоматически. Балки можно удалять и копировать;

нанесение групп плит осуществляется путем выбора в меню нужной марки плит и их вертикальности, указанием в чертеже положений крайних плит (при перемещениях курсора прорисовываются плиты строящейся группы, и подсвечивается их количество) и затем уточнением их параметров. Группу плит можно "образмерить", т.е. будут проставлены толщины всех плит группы;

Генерация разреза осуществляется поэтапным заданием следующих сведений:

число этажей в разрезе, наличие фундамента и покрытия;

для каждого из них задаются соответствующие планы путем указания в чертеже модуля "План строительной подосновы" с комплектом параметров плана. Программа не позволит указать что-либо другое. Для этажей, начиная со 2-го, можно выбрать план перекрытия, которое будет являться полом этого этажа;

для всех этажей вводятся уровни пола, плюс уровень подошвы фундамента (если есть фундамент) плюс уровень низа покрытия (потолка, если есть покрытие) или верха последнего этажа (если нет покрытия). Заданные уровни при генерации разреза наносятся на чертеж как отметки высоты;

на одном из относящихся к разрезу планов указывается секущая - ломаная вдоль осей X и Y, буквенное обозначение и масштаб разреза;

генерируется сам чертеж разреза.

Для наглядного анализа в процессе проектирования подосновы предусмотрен просмотр разрезов с пошаговыми смещением и вращением секущей плоскости в пространстве.

Колонны, перегородки и другие элементы строительной подосновы при помещении в чертеж автоматически объявляются компоновочными блоками для последующей компоновки оборудования. ПП плана подосновы запоминается в чертеже в модуле "План строительной подосновы", видимая его часть включает два отрезка начальных осей X и Y (с их маркировкой) и два размера (если были проставлены).

Изложенные в настоящей работе параметрическое представление, состав и особенности реализации операций проектирования являются моделью чертежей строительной подосновы - общей составляющей большого числа чертежей различных марок системы проектной документации для строительства, повышающей эффективность систем автоматизированного проектирования, особенно при проектировании реконструкции предприятий. Ключевым моментом как в реализации параметрического представления, так и в обеспечении привычного графического пользовательского интерфейса явилась эксплуатация модулей в чертеже, объединяющих видимую графическую часть и невидимые параметры.

За несколько лет эксплуатации подсистемы проектирования строительной подосновы в условиях проектно-конструкторского отдела крупного химического предприятия разработаны несколько сот чертежей строительной подосновы, что подтверждает эффективность и практическую значимость изложенных подходов к моделированию чертежей строительной подосновы.

## Литература